\documentclass[a4paper,12pt]{article}
\usepackage{amssymb}

\begin{document}

{\centerline {\bf Notes on Unfair Papers by Mebarki et al.}}
{\centerline {\bf on ``Quantum Nonsymmetric Gravity''}}

\vskip20pt

{\centerline {Noboru Nakanishi\footnote{Professor Emeritus of Kyoto
University
}}

{\centerline {\small 12-20 Asahigaoka-cho, Hirakata, 573-0026, Japan}}

{\centerline {and}}

{\centerline {Izumi Ojima}}

{\centerline {\small Research Institute for Mathematical Sciences,}

{\centerline {\small Kyoto University, Kyoto, 606-8502, Japan}}

\vskip15pt

\noindent {\small It is pointed out that the essential parts of some recent 
papers by Mebarki {\it et al.} (hep-th/9911045, hep-th/9911046, 
hep-th/9911048, hep-th/9911049, dated 6 Nov.1999) are taken from 
a book written by Nakanishi and Ojima, published in 1990.}

\vskip20pt

\noindent The following four papers, which have appeared as E-prints very
recently, have come to our attention:

\vskip5pt

\noindent [MMZ] N. Mebarki, A. Maireche and S. Zaim, $N = 1,\ D = 4$ Quantum
Nonsymmetric supergravity (hep-th/9911045),

\noindent [MMBB] N. Mebarki, A. Maireche, A. Boudine and A. Benslama,
Symmetries
of Quantum Nonsymmetric Gravity (hep-th/9911046),

\noindent [MM] N. Mebarki and A. Maireche, Quantum Nonsymmetric Gravity
Geometric Commutation Relations (hep-th/9911048),

\noindent [MMH] N. Mebarki, A. Maireche and M. Houchine, QNGT Sixteen
Dimensional GL(4,R)-like Superalgebra (hep-th/9911049).

\vskip5pt
Taking advantage of the situation that nobody has discussed the
quantum theory of {\it nonsymmetric gravity}, these authors pretend as if
all of these papers were their original work. However, apart from some
considerations at the Lagrangian level, the essential parts are taken from
Chapter 5 of the following book published in 1990 {\it with absolutely no
mention to it}:

\vskip5pt

\noindent [NO]  N. Nakanishi and I. Ojima, {\it Covariant Operator Formalism
of Gauge Theories and Quantum Gravity} (World Scientific, Singapore, 1990).

\vskip5pt

\noindent Moreover, Mebarki {\it et al.} delete all the references to
the original papers on quantum Einstein gravity by Nakanishi (partly with
some collaborators), published in 1978-1985.

\vskip5pt
{\it What they reproduce from the book [NO] are not merely contents but also
bulk of sentences}, though, of course, with straightforward modifications
necessary for pretending as if those were their own sentences. As a result,
it becomes very easy to identify which parts are taken by them. As for [MMZ]
and [MMBB], see Appendix.

Aside from Introduction and Conclusion, {\it the whole papers of} [MM]
{\it and} [MMH] {\it are taken almost faithfully} (with some portions
skipped and with some mistakes committed) {\it from Section 5.6 (pp.319-333)
of} [NO] {\it and from Section 5.5 (pp.312-318) of} [NO], {\it
respectively}. Mebarki {\it et al}. have copied even simple misprints of a
mathematical expression in [NO], which could easily have been corrected
if they had checked mathematical formulae.

The geometric commutation relation and the sixteen dimensional
superalgebra  or choral symmetry (named by Nakanishi) are very
remarkable results {\it characteristic to quantum Einstein gravity}. To
derive them, one must make full use of the equal-time commutation
relations (between field operators and their time derivatives), whose
calculations are extremely elaborate. Mebarki et al. write merely canonical
commutation relations but one can find no evidence that they have actually
calculated the equal-time commutation relations (between field operators and
their time derivatives) of quantum nonsymmetric gravity. These expressions
must be {\it different} from those of quantum Einstein gravity. Therefore
there
is no reason to believe that the geometric commutation relation and the
sixteen dimensional superalgebra remain valid in quantum nonsymmetric
gravity.

\vskip10pt
We hope that Mebarki, Maireche and their collaborators refrain from
the unfair conduct shameful as scientists.

\vskip15pt

\section*{Appendix}
\begin{itemize}
\item Paper[MMZ]:
\end{itemize}
\begin{description}
 \item[i)] P.9, line 7 $\--$ p.10, line 6, is taken from p.300 (Chapter 5)
 of [NO].
 \item[ii)] p.11, line 17 $\--$ p. 12, line 5, is from p.340 (Chapter 5)
 of [NO].
\end{description}

\begin{itemize}
\item Paper[MMBB]:
\end{itemize}
\begin{description}
 \item[i)] The opening (ten lines on p.2) is a faithful copy of the opening
of Chapter 5 on p.281 of [NO].
 \item[] One can easily find more such as
 \item[ii)] p.5, line 23 $\--$ p.6, line 11, is an imitation of p.158
(Chapter 3) of [NO] [with an erroneous change of ``confine'' into
``confirm''].
 \item[iii)] p.6, line 16 $\--$ line 19 is taken from p.14 (Chapter 1) of
[NO].
 \item[iv)] p.7, line 9 $\--$ line 13, is borrowed from p.304 (Chapter 5)
 of [NO]; as pointed out in the text, however, it is quite unlikely that the
 sentence ``all the equal time commutators can be obtained in closed form''
 is realized in the framework of quantum nonsymmetric gravity.
 \item[v)] p.9, line 1 $\--$ line 4, is taken from p.159 (Chapter 3)
 of [NO].
\end{description}
\end{document}